\documentclass[preprint2]{aastex}
\usepackage{textcomp}
\usepackage{url}

\slugcomment{To appear in ApJ. Letters.}

\shorttitle{Modeling Shocks at Voyager 1}
\shortauthors{Kim et al.}

\begin{document}



\title{MODELING SHOCKS DETECTED BY \textit{VOYAGER 1} IN THE LOCAL INTERSTELLAR MEDIUM}


\author{T. K. Kim\altaffilmark{1}, N. V. Pogorelov\altaffilmark{1,2}, and L. F. Burlaga\altaffilmark{3}}




\altaffiltext{1}{Center for Space Plasma and Aeronomic Research, University of Alabama in Huntsville, Huntsville, AL 35805, USA}
\altaffiltext{2}{Department of Space Science, University of Alabama in Huntsville, Huntsville, AL 35805, USA}
\altaffiltext{3}{NASA Goddard Space Flight Center, Code 673, Greenbelt, MD 20771, USA}

\begin{abstract}
The magnetometer (MAG) on \textit{Voyager 1} (\textit{V1}) has been sampling the interstellar magnetic field (ISMF) since August 2012. The \textit{V1} MAG observations have shown draped ISMF in the very local interstellar medium disturbed occasionally by significant enhancements in magnetic field strength. Using a three-dimensional, data driven, multi-fluid model, we investigated these magnetic field enhancements beyond the heliopause that are supposedly associated with solar transients. To introduce time-dependent effects at the inner boundary at 1 astronomical unit, we used daily averages of the solar wind parameters from the OMNI data set. The model ISMF strength, direction, and proton number density are compared with \textit{V1} data beyond the heliopause. The model reproduced the large-scale fluctuations between 2012.652 and 2016.652, including major events around 2012.9 and 2014.6. The model also predicts shocks arriving at \textit{V1} around 2017.395 and 2019.502. Another model driven by OMNI data with interplanetary coronal mass ejections (ICMEs) removed at the inner boundary suggests that ICMEs may play a significant role in the propagation of shocks into the interstellar medium.
\end{abstract}

\keywords{ISM: magnetic fields --- Sun: heliosphere --- solar wind --- methods: numerical --- magnetohydrodynamics (MHD)}

\section{Introduction}

The \textit{Voyager 1} (\textit{V1}) spacecraft crossed the heliopause into the local interstellar medium (LISM) at 122 astronomical units (au) in August 2012. This major milestone was preceded by a two-step increase in galactic cosmic ray flux accompanied by a significant drop in anomalous cosmic ray intensities leading to the event, and confirmed by the high electron densities observed by the plasma wave science (PWS) instrument in April--May 2013 \citep{Burlaga2013Science,Stone2013Science,Gurnett2013Science}. The magnetometer (MAG) aboard \textit{V1} has observed consistently large magnetic field strength above 0.4 nT in the very local interstellar medium (VLISM) near the heliopause, which was considerably stronger than any previously measured values in the inner heliosheath \citep{Burlaga2014ApJ,BurlagaNess2014ApJL,BurlagaNess2016ApJ}. The interstellar magnetic field (ISMF) exhibited compressible, weakly turbulent fluctuations while the azimuthal angle $\lambda$ and the elevation angle $\delta$ were observed to increase/decrease linearly. The direction of the ISMF at \textit{V1} is significantly different from the Parker spiral direction, indicating a draped ISMF in the VLISM \citep{Burlaga2014ApJ,BurlagaNess2014ApJL,Burlaga2015ApJL,BurlagaNess2016ApJ}.

Since 2012.65 (DOY 238), \textit{V1} MAG recorded two significant jumps in ISMF strength around 2012.92 (DOY 336) and 2014.65 (DOY 236). Both of these disturbances were preceded by electron plasma oscillation events which are strong indicators of shocks \citep{Gurnett2015ApJ}, and followed by relatively quiet periods that ended with abrupt decreases in ISMF strength in 2013.35 (DOY 130) and 2015.37 (DOY 136) \citep{BurlagaNess2016ApJ}. Additionally, \textit{V1} PWS observed an intense plasma oscillation event in April--May 2013, though MAG did not measure a meaningful jump in ISMF strength following the event unlike in 2012 and 2014. It is clear from these observations that \textit{V1} is still immersed in a region disturbed by shocks and pressure pulses of solar origin. First suggested by \cite{Gurnett1993Science}, the idea of heliospheric shocks propagating across the heliopause into the LISM has been supported by a number of models \citep{WhangBurlaga1994JGR,ZankMuller2003JGR,Washimi2011MNRAS,Washimi2015ApJ,PogorelovZank2005ESASP,Pogorelov2012AIP}.

There have been attempts to model shocks beyond the heliopause using near-Earth solar wind (SW) data. \cite{Liu2014ApJL} investigated one-dimensional (1-D) propagation of interplanetary coronal mass ejections (ICMEs) and associated shocks from 1 to 120 au using a magnetohydrodynamics (MHD) model and the \textit{Wind} spacecraft data. The 1-D MHD model showed formation of a large merged interaction region (MIR) from a series of ICMEs encountered at 1 au by \textit{Wind} in March 2012. The model included the effects of pickup ions, but neglected transition across the termination shock to the inner heliosheath, and also across the heliopause to the LISM. To account for the large uncertainties of shock propagation through the inner heliosheath and VLISM which were not included in the model, \cite{Liu2014ApJL} had to make ad hoc adjustments to the shock arrival time at 120 au using shock passage through the Earth's bow shock and the magnetosheath as an analogue.

More recently, \cite{Fermo2015JPhCS} used a fully three-dimensional (3D) multi-fluid MHD-neutral model to simulate shocks in the LISM, which had the advantage of including heliospheric structures lacking in 1-D models. The \cite{Fermo2015JPhCS} model assumed spherical symmetry at 1 au where hourly averaged OMNI data (spacecraft-interspersed, near-Earth SW data available at \url{https://omniweb.gsfc.nasa.gov/}) were used as inner boundary conditions. Although the model showed magnetic field and density enhancements related to global MIRs (GMIRs) propagating across the heliopause, the location of the heliopause was incorrect by 20--30 au, which made it difficult to directly assess each individual event observed by \textit{V1}. The main objectives of our study are as follow: (1) to attribute the shocks observed by \textit{V1} in the LISM to SW observations in the OMNI database at 1 au and (2) to identify the relative contributions of ICMEs and corotating interaction regions (CIRs) to the modeled shocks.

\section{Model}

We devised a 3D multi-fluid model within the framework of Multi-scale Fluid-kinetic Simulation Suite (MS-FLUKSS) to simulate the interaction between the SW and the partially ionized LISM \citep{Borovikov2013ASP,Pogorelov2014XSEDE}. The model consists of five separate fluids: one plasma fluid and four populations of neutral hydrogen atoms originating in different regions - i.e., in the undisturbed LISM, VLISM around the heliopause, inner heliosheath, and the super-Alv\'{e}nic SW. While the plasma flow is governed by ideal MHD equations, the flows of neutral hydrogen atoms are described hydrodynamically by means of multi-component Euler equations \citep{Zank1996JGR,Pogorelov2006ApJ}. We make a simplifying assumption that pickup ions resulting from the charge-exchange process between ions and neutral atoms are immediately equilibrated with thermal ions to form an isotropic mixture.

For computational efficiency, we divide the time-dependent simulation into two parts. In the first part, the SW is propagated from 1 to 12 au using a base grid of 256$\times$128$\times$64 cells in a spherical coordinate system ($r$, $\phi$, $\theta$). Subsequently, we use the time-dependent solutions saved at 12 au as inner boundary conditions for the second part where we employ a base grid of 640$\times$128$\times$64 cells with the outer boundary defined at 1000 AU. In both parts, the radial grid size varies with heliocentric distance such that $\Delta$r is approximately 0.03 au at 1 au, 0.13 au at 12 au, 0.4 au at 120 au, and 20 au at 1000 au, for example. In the non-radial directions, the base grid $\Delta\phi$ and $\Delta\theta$ are both $\sim$2.8\textdegree. The base grid is too coarse to resolve shocks at large distances, so adaptive mesh refinement (AMR) technique is used to selectively refine the grid in the inner heliosheath and VLISM, particularly around the heliopause. With AMR, the radial grid size is reduced to $\Delta$r = 0.036 au at 80 au, 0.043 au at 100 au, 0.025 au at 120 au, and 0.045 au at 140 au, whereas the non-radial grid size becomes as little as 0.18\textdegree\ at 120 au. 

\begin{figure}[h]
\begin{center}
\noindent\includegraphics[width=0.5\textwidth, angle=0]{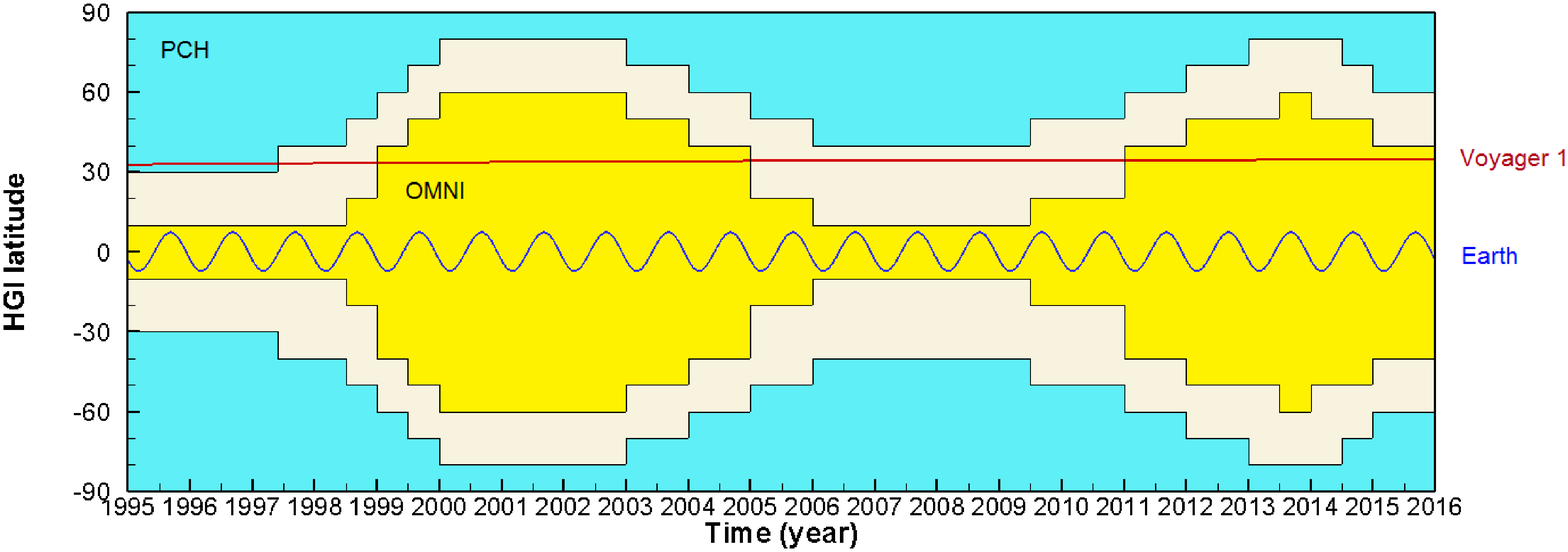}
\noindent\includegraphics[width=0.5\textwidth, angle=0]{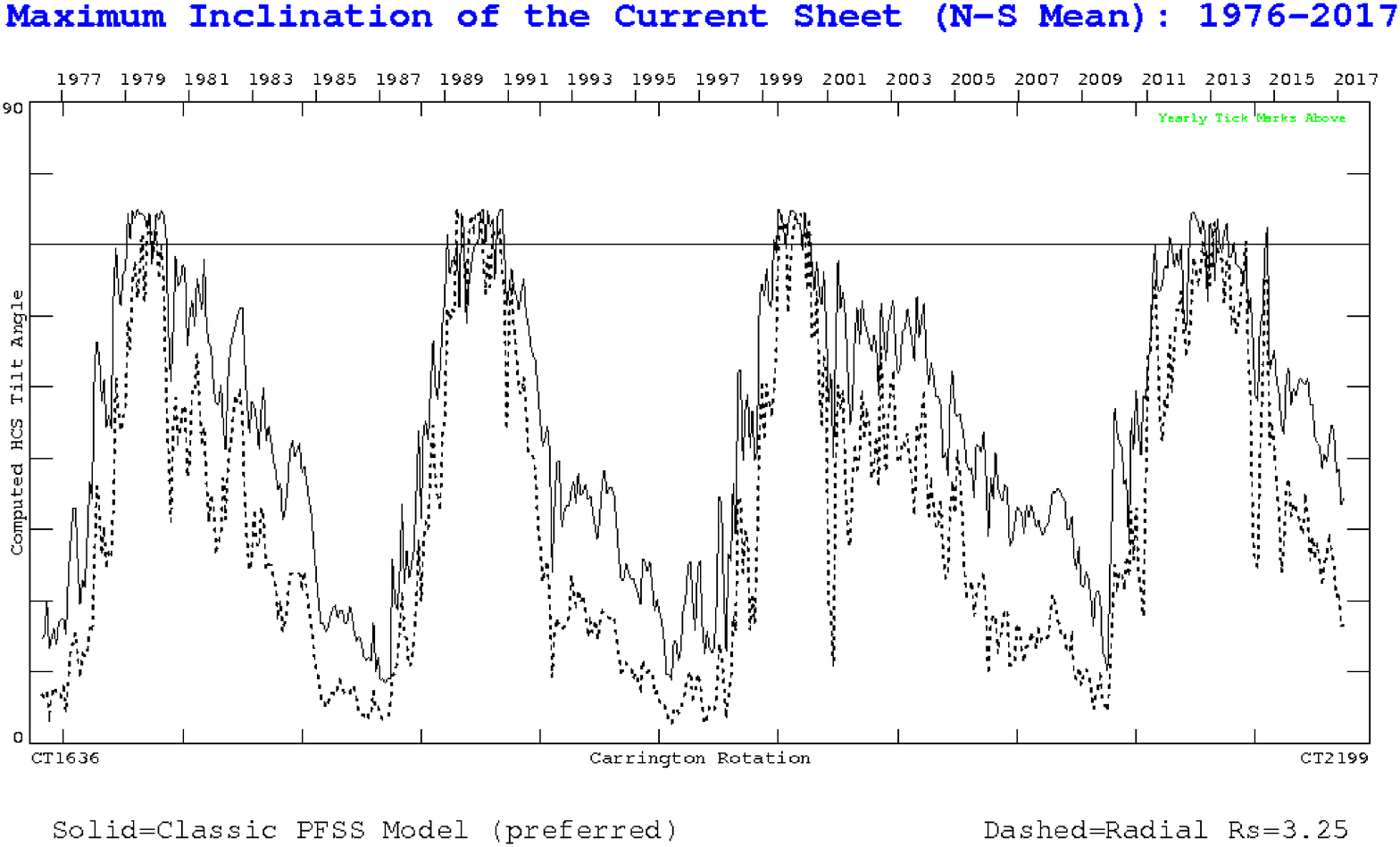}
\end{center}
\caption{Top: A diagram showing the temporal variation of the latitudinal extents of the PCHs (light blue) and OMNI data (yellow) at 1 AU. Also shown are the heliographic latitudes of Earth (blue) and Voyager 1 (red). Bottom: Average HCS tilt shown as a function of time (courtesy of WSO).}
\label{fig1}
\end{figure}

We used OMNI daily averaged plasma and interplanetary magnetic field data to introduce time-dependent effects at 1 au. The inner boundary frame is divided into different regions filled with OMNI data and idealized polar coronal hole (PCH) values whose latitudinal extents vary with time as illustrated in the top panel of Figure \ref{fig1}. In the equatorial OMNI region of each daily frame at 1 au, we filled the 360\textdegree\ longitudinal space centered around Earth using 27 days of data (up to 13 days from the past/future) with a simplifying assumption that the SW propagated radially outward with a spiral magnetic field at 1 au. We estimated the SW parameters in PCHs using empirical correlations \citep{Pogorelov2013ApJ} to best fit the Ulysses data at high heliographic latitudes. The interface between OMNI and PCH regions is 20\textdegree--30\textdegree\ wide and are linearly interpolated over. The procedure is described in more detail by \cite{Kim2016ApJ} who used the same boundary conditions to reproduce large-scale fluctuations of the SW properties at \textit{Ulysses}, \textit{Voyager}, and \textit{New Horizons} at various distances and latitudes between 1 and 80 au.

While we estimated the magnetic field components at 1 au from OMNI $|\textbf{B}|$ data in the form of Parker spiral field as done by \cite{Kim2016ApJ}, we further introduced a heliospheric current sheet (HCS) in the form of a tilted circle \citep{Pogorelov2007ApJ} whose tilt with respect to the Sun's rotation axis changed as a function of time according to the average HCS tilt provided by Wilcox Solar Observatory (WSO) (see the bottom panel of Figure \ref{fig1}). Furthermore, the polarity of magnetic field above and below the HCS at 1 au was instantaneously and simultaneously reversed during solar maximum in 2000 and 2013. In reality, polarity reversal is more complicated and occurs gradually over several months.

At the outer boundary at 1000 au, we set the inflow speed, direction, and temperature of interstellar hydrogen to 25.4 km s$^{-1}$, 75.7\textdegree\ ecliptic inflow longitude, -5.1\textdegree\ ecliptic inflow latitude, and 7500 K, respectively, as suggested by \cite{McComas2015ApJS} based on observations by \textit{Interstellar Boundary Explorer} (\textit{IBEX}). We also use interstellar proton and hydrogen atom densities of 0.09 cm$^{-3}$ and 0.154 cm$^{-3}$ as well as ISMF strength and direction of 3 $\mu$G, 226.99\textdegree\ ecliptic longitude, and 34.82\textdegree\ ecliptic latitude which produced the best model fit to the ``ribbon'' of intense energetic neutral atom emissions observed by \textit{IBEX} \citep{Zirnstein2016ApJL}.




\section{Results}

\begin{figure}[h]
\begin{center}
\noindent\includegraphics[width=0.47\textwidth, angle=0]{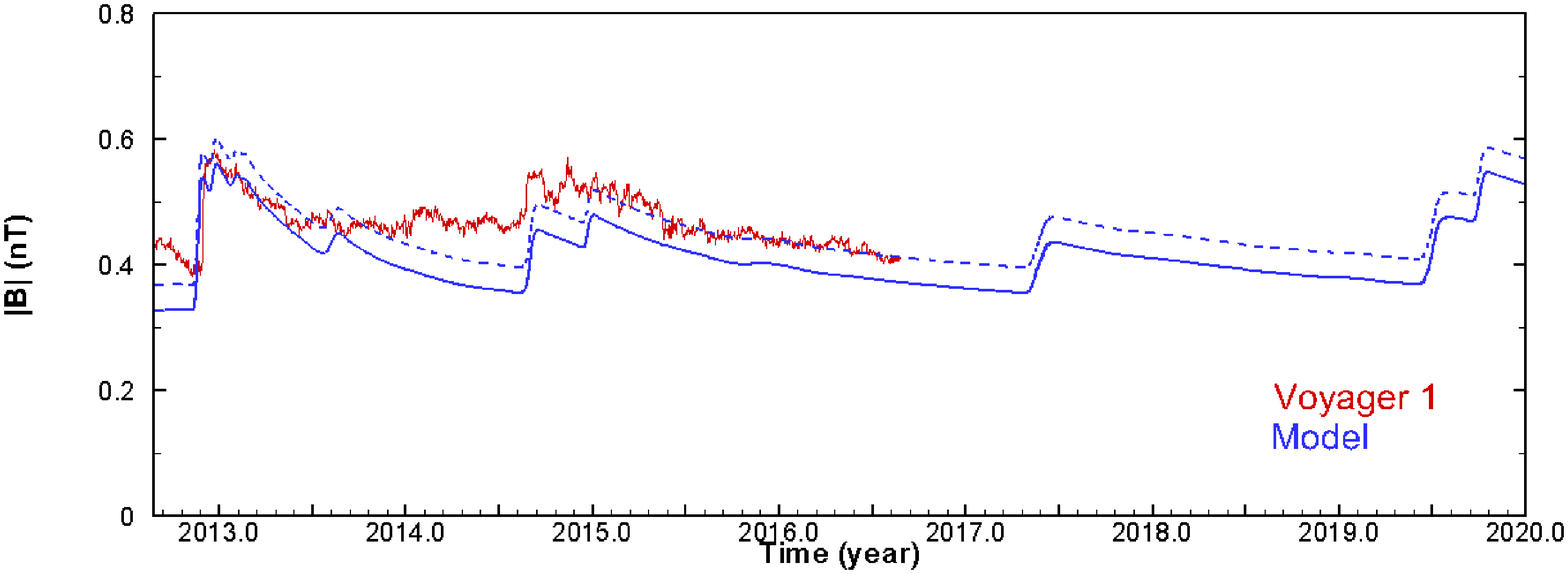}
\noindent\includegraphics[width=0.47\textwidth, angle=0]{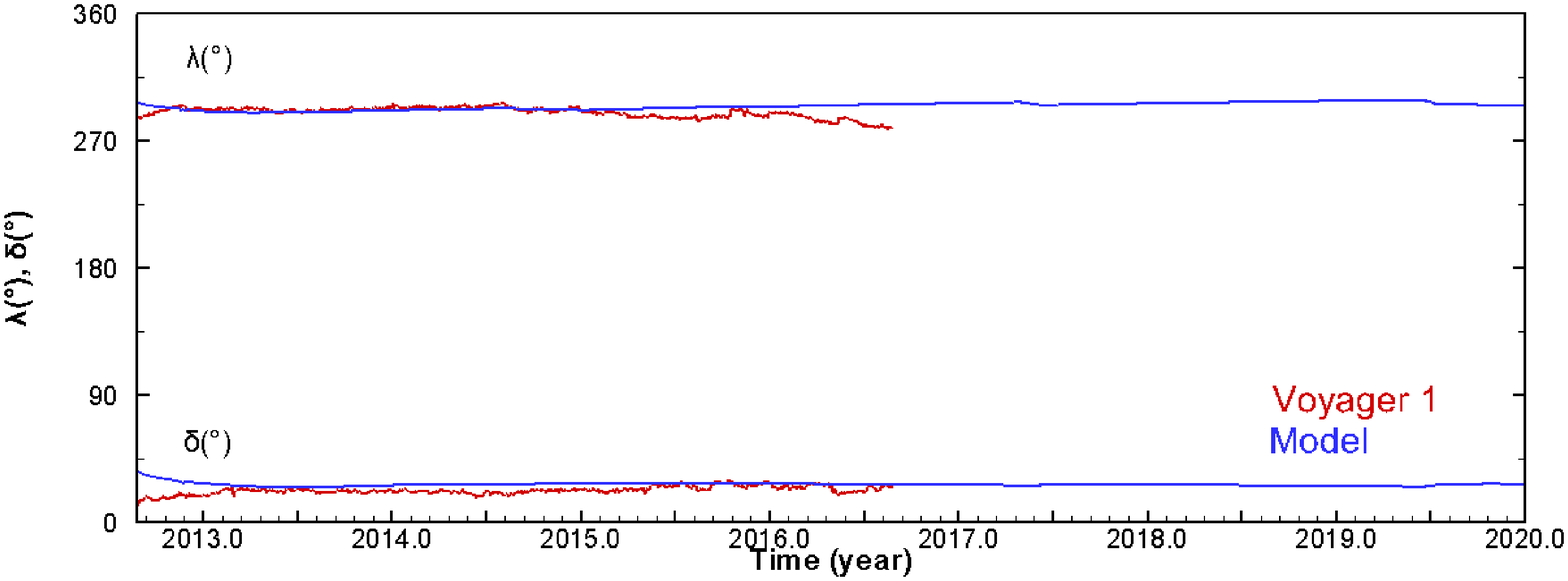}
\noindent\includegraphics[width=0.47\textwidth, angle=0]{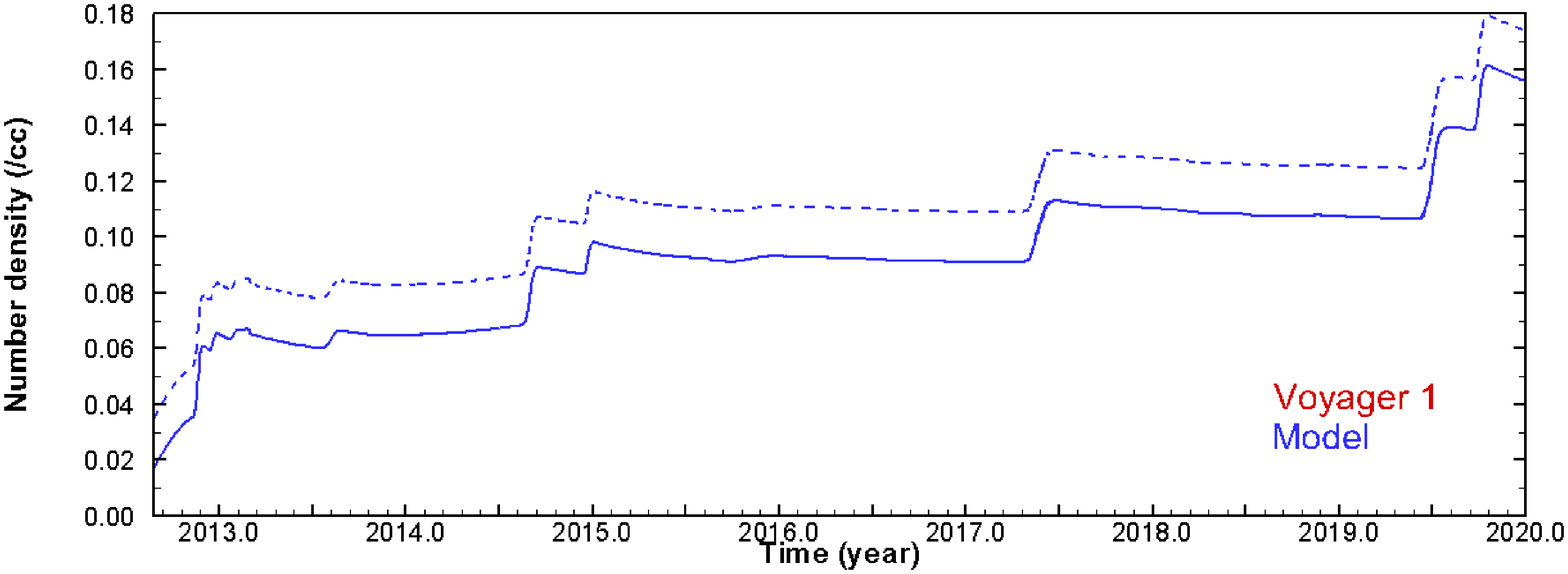}
\noindent\includegraphics[width=0.47\textwidth, angle=0]{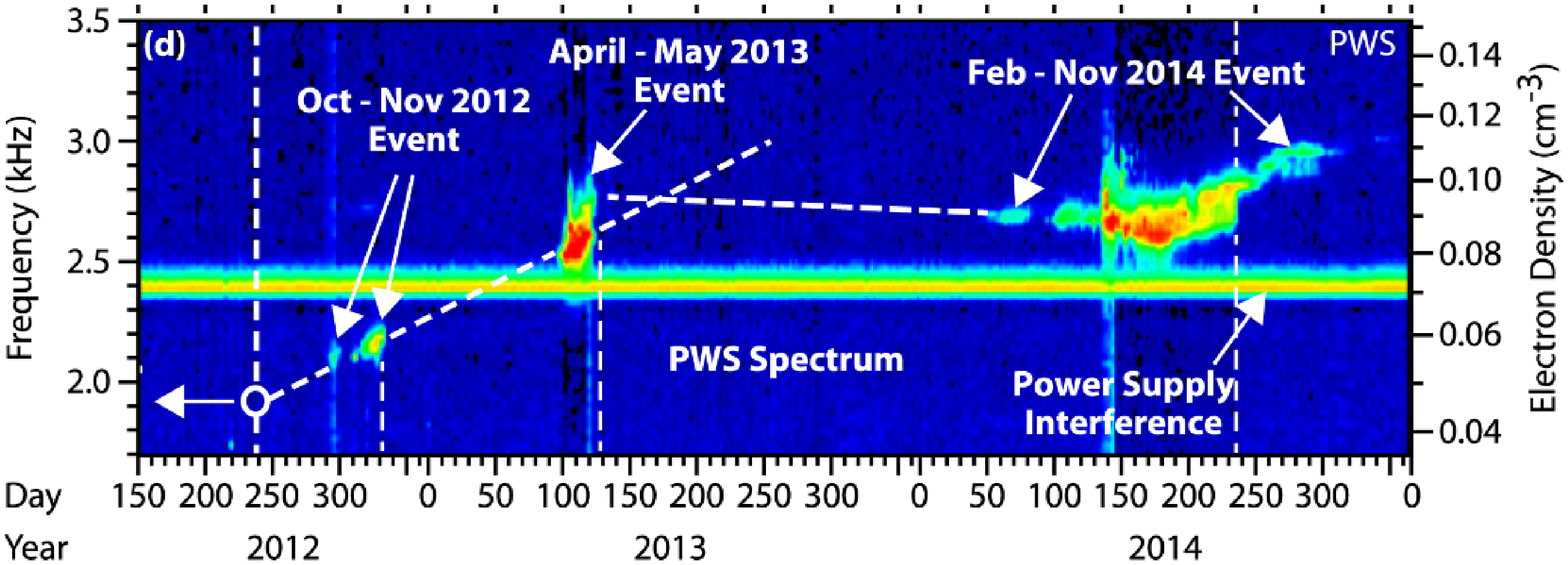}
\end{center}
\caption{Model $|\textbf{B}|$ and the azimuth and elevation angles $\lambda$ and $\delta$ are shown in blue compared to the daily averaged \textit{V1} MAG data which are shown in red. The model proton number density is shown in blue compared with \textit{V1} PWS observations taken from \cite{Gurnett2015ApJ} with permission of the AAS. The model $|\textbf{B}|$ and density are shifted up by 0.04 nT and 0.018 cm$^{-3}$ (dashed blue) to best match the \textit{V1} MAG data during the undisturbed period in 2016 and electron plasma densities derived from PWS observations, respectively.}
\label{V1}
\end{figure}

The top two panels of Figure \ref{V1} show the model $|\textbf{B}|$, the azimuthal angle $\lambda$, and the elevation angle $\delta$ compared with \textit{V1} MAG data in RTN coordinates between 2012.652 and 2016.652, though the model is shown extended out to 2020.0 until the last shock generated by the time-dependent boundary conditions reaches \textit{V1}. Fluctuations of the model interstellar proton number density are also shown along with a spectrogram of the wideband electric field spectral densities measured by the \textit{V1} PWS instrument \citep{Gurnett2015ApJ} in the two bottom panels of Figure \ref{V1}. The plasma science instrument on \textit{V1} is not functioning, but it is still possible to derive density from electron plasma oscillation events detected by PWS assuming the observed electron plasma frequency $f_{p}$ = 8980$\sqrt{n_{e}}$ Hz, where $n_{e}$ is in cm$^{-3}$ \citep{Gurnett2015ApJ}.

There is a difference of 0.06--0.11 nT between the model $|\textbf{B}|$ and \textit{V1} MAG data immediately after the heliopause crossing from 2012.652 to 2012.880. The observed $|\textbf{B}|$ decreased from 0.44 to 0.39 nT during this interval, but the model $|\textbf{B}|$ remained steady around 0.330 nT while the model density smoothly increased from 0.0172 to 0.0360 cm$^{-3}$. The azimuthal and elevation angles $\lambda$ and $\delta$ of the model $|\textbf{B}|$ changed from 297\textdegree\ to 292\textdegree\ and from 36\textdegree\ to 28\textdegree, respectively, whereas the observed $\lambda$ and $\delta$ increased from 286\textdegree\ to 295\textdegree\ and from 16\textdegree\ to 20\textdegree. The initial discrepancy of 11\textdegree\ (20\textdegree) between the model and observed $\lambda$ ($\delta$) decreased to 3\textdegree\ (8\textdegree) at 2012.880 and remained within 16\textdegree\ (8\textdegree) until 2016.650. The observed decrease in $|\textbf{B}|$ may be explained by a heliospheric boundary layer resulting from draping of the ISMF around the heliopause \citep{Pogorelov2017ApJ}. The heliospheric boundary layer is characterized by increasing density on the LISM side of the heliopause, which is clearly reproduced by this model. The initial discrepancy between the model and observed $|\textbf{B}|$, $\lambda$ and $\delta$ in the vicinity of the heliopause may be associated with uncertainties in the heliopause motion due to time-dependent effects and plasma instability at the heliopause (e.g., \cite{BorovikovPogorelov2014ApJL}).

The first modeled shock arrived at \textit{V1} around 2012.890 marked by step-like increases in $|\textbf{B}|$ from 0.330 to 0.537 nT and density from 0.0360 to 0.0607 cm$^{-3}$. The ratio of the model $|\textbf{B}|$ across the shock $B_{2}/B_{1}$ is 1.63 which is 11\% larger than the observed ratio of 1.47 \citep{BurlagaNess2016ApJ}. Considering the uncertainties of the observations, these ratios agree favorably. Following the shock passage, the model $|\textbf{B}|$ decreased smoothly until the middle of 2013, in general agreement with observations. However, the model did not reproduce the sharp decrease in the observed $|\textbf{B}|$ at 2013.353 while showing a moderate bump around 2013.600 which was not observed by \textit{V1} MAG. \cite{BurlagaNess2016ApJ} suggest two possibilities for the abrupt decrease in $|\textbf{B}|$: stationary current sheets embedded in the LISM plasma or a reverse shock / pressure waves. In the first case, we do not expect our model to reproduce this structure because such structures are not included in the model, but it may be possible in the second case with a more refined grid. We point out that the abrupt decrease at 2013.353 occurred over 2--3 days during which V1 moved away from the Sun by 0.02--0.03 au. However, the radial grid size of the model at \textit{V1} at that time was 0.0256 au, which might not be sufficiently small to resolve such a narrow, small-scale structure (i.e., $B_{2}/B_{1}$ = 1.07).

The abrupt decrease in the observed $|\textbf{B}|$ was followed by a relatively quiet interval from 2013.362 to 2014.641 characterized by small amplitude fluctuations in $|\textbf{B}|$ having a Kolmogorov spectrum \citep{Burlaga2015ApJL}. In contrast, the model $|\textbf{B}|$ increased by $\sim$0.03 nT around 2013.600 and decreased almost linearly down to 0.358 nT until mid-2014. While we did not attempt to track shocks driven by each individual ICMEs in the model, we speculate that the moderate increase in the model $|\textbf{B}|$ at 2013.600 may have been triggered by a pressure wave delivered by a structure formed primarily by ICMEs in the OMNI data with large longitudinal separation from \textit{V1}. We further point out the systematically lower model $|\textbf{B}|$ and density in comparison to observations during this quiet interval. The pattern of lower-than-observed model values which persisted until at least mid-2016 may be associated with a large rarefaction region trailing the GMIR that drove the 2012 shock in the model. When the rarefaction region reached the heliopause in the model, it caused the heliopause to oscillate, and a significant rarefaction region developed behind the heliopause and propagated into the LISM. Observations do not suggest any dramatic movement of the heliopause or such large decrease of the ISMF after the 2012 shock. The modeled shocks behind the heliopause are consistently stronger than observed, so we suspect that GMIRs and associated structures in the model may have been somewhat exaggerated in their scale and influence on the heliopause.

The end of the quiet interval was marked by the second modeled shock that arrived at \textit{V1} around 2014.665 when the model $|\textbf{B}|$ and density jumped from 0.358 to 0.455 nT and from 0.0690 to 0.0891 cm$^{-3}$, respectively. The arrival time of the modeled shock closely matches that of the observed shock at 2014.648, but the ratios of the model $|\textbf{B}|$ (1.27) and density (1.29) are 12\% and 16\% higher than the observed values of 1.13 and 1.11 \citep{Gurnett2015ApJ,BurlagaNess2016ApJ}, though we estimate the uncertainty of the observed ratios to be $\sim$5\%. Shortly after the passage of the second shock, the model showed another step-like enhancement in $|\textbf{B}|$ and density at $\sim$2014.978. The model $|\textbf{B}|$ steadily decreased during the relatively undisturbed period afterwards, in agreement with observations.

\cite{BurlagaNess2016ApJ} reported a linear decrease in the observed $|\textbf{B}|$ from 2014.833 to 2015.370 ending with an abrupt decrease from 0.494 to 0.456. \cite{BurlagaNess2016ApJ} also pointed out a 28-day quasi-period oscillation of $|\textbf{B}|$ during this interval, which was followed by a quiet period of steady decline in the observed $|\textbf{B}|$. The model reproduced the steady decrease in $|\textbf{B}|$ from late-2014 to mid-2016. We offer comparison between the model and observations until 2016.652 because \textit{V1} MAG data are only available up to that date at the moment. However, we extend the model out to 2020.0 until every shock generated by the time-varying boundary conditions propagates out to \textit{V1}.

The third modeled shock arrives at \textit{V1} around 2017.395 when $|\textbf{B}|$ and density increase from 0.357 to 0.435 nT and from 0.0913 to 0.113 cm$^{-3}$. The ratios of the model $|\textbf{B}|$ and density across the shock are 1.22 and 1.24, respectively. The final shock generated by the time-dependent boundary conditions arrives at \textit{V1} around 2019.502 when $|\textbf{B}|$ and density increase from 0.369 to 0.474 nT and from 0.107 to 0.139 cm$^{-3}$ with $|\textbf{B}|$ and density ratios of 1.28 and 1.30 across the shock. Similar to the mid-2014 shock, the model shows another step-like increase in $|\textbf{B}|$ from 0.470 to 0.545 nT and density from 0.138 to 0.161 cm$^{-3}$ around 2019.754 shortly after the shock passage. The widths of these shocks appear broader than the previous shocks due to the relatively large grid size at larger distances. The first two shocks passed \textit{V1} at 122.44 and 128.80 au where the radial grid sizes are 0.0252 and 0.0262 au, respectively. When the latter two shocks reach \textit{V1} at 138.55 and 146.07 au, the radial grid sizes are 0.0427 and 0.0528 au, respectively.

\begin{figure}[h]
\begin{center}
\noindent\includegraphics[width=0.47\textwidth, angle=0]{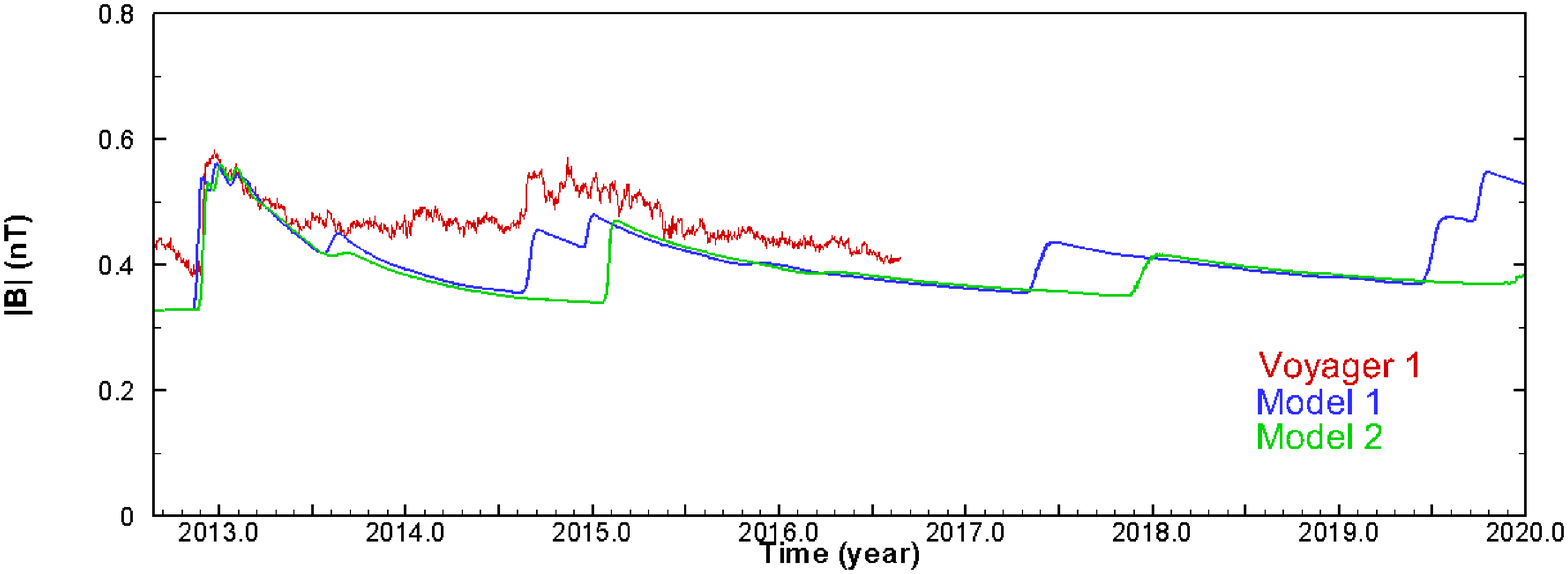}
\noindent\includegraphics[width=0.47\textwidth, angle=0]{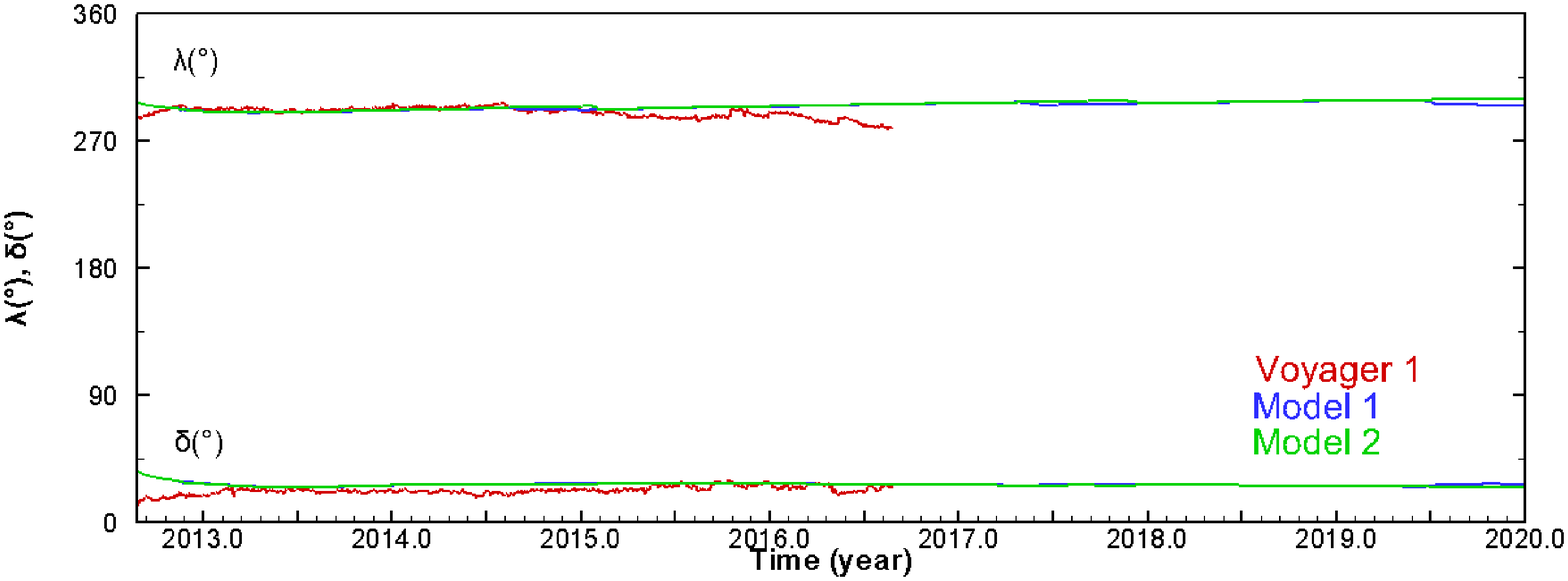}
\noindent\includegraphics[width=0.47\textwidth, angle=0]{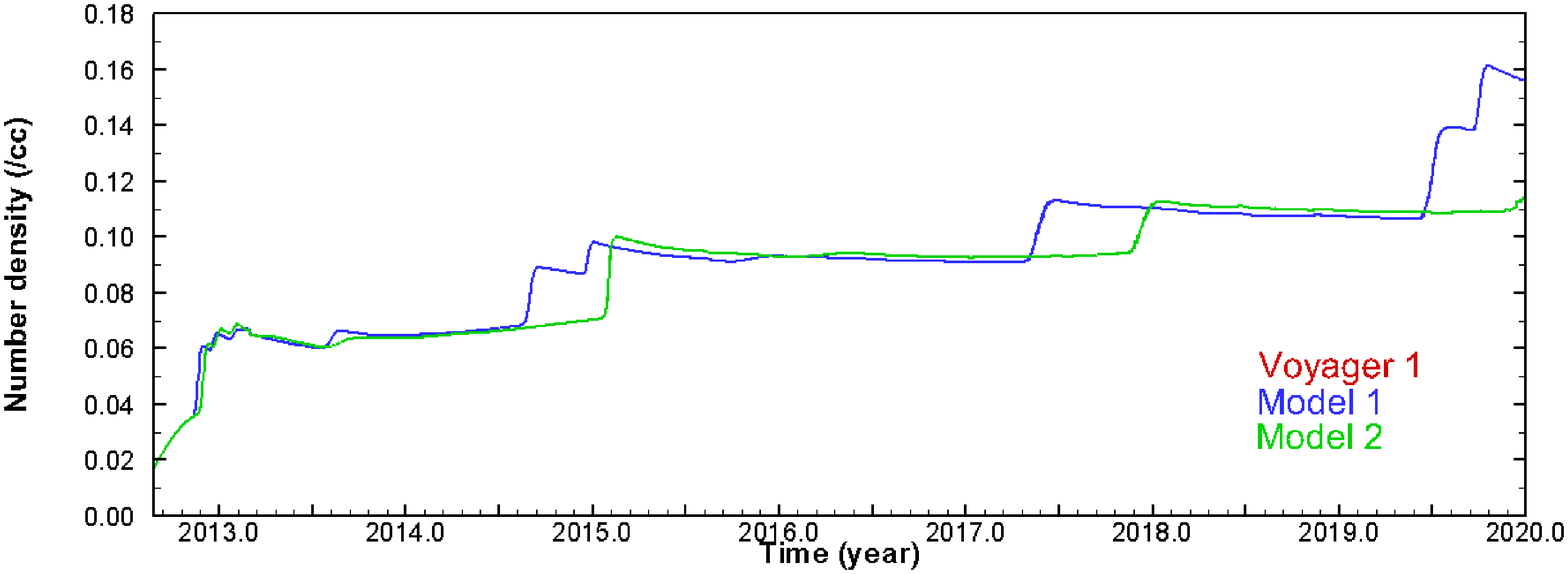}
\end{center}
\caption{$|\textbf{B}|$, $\lambda$, $\delta$, and proton number densities for Model 1 (including all OMNI data) and Model 2 (ICMEs excluded from OMNI data) are shown in blue and green, respectively. The daily averaged \textit{V1} MAG data are shown in red.}
\label{V1noICMEs}
\end{figure}

It is interesting to see how much ICMEs affect the shocks modeled at \textit{V1} beyond the heliopause. To estimate the contribution of ICMEs, we used the ICME lists for \textit{ACE} (\url{http://www.srl.caltech.edu/ACE/ASC/DATA/level3/icmetable2.htm}) and \textit{Wind} (\url{https://wind.nasa.gov/ICMEindex.php}) as reference to identify and remove ICMEs at 1 au, and the resulting data gaps were linearly interpolated. Thus, the boundary conditions at 1 au in this case would only consist of ambient and corotating streams. We followed the same procedure described in the previous section to perform a time-dependent simulation with these boundary conditions. The results are labeled as Model 2 in Figure \ref{V1noICMEs} where Model 1 refers to the original results shown in Figure \ref{V1}.

The arrival time of the late-2012 shock in Model 2 is around 2012.916, and the $|\textbf{B}|$ and density ratios across this shock are 1.62 and 1.68, respectively. These values are nearly the same as in Model 1, suggesting that this shock may have been driven by an MIR consisting primarily of corotating interaction regions (CIRs) in both models. Contrastingly, the second shock that reached \textit{V1} at 2014.665 in Model 1 arrived significantly later in Model 2 at 2015.090. The $|\textbf{B}|$ and density ratios across this shock are 1.38 and 1.40, respectively, which are slightly larger than the ratios in Model 1 by 8.7\% and 8.5\%. Comparison of the models with observations suggests that the mid-2014 shock may have been driven by an MIR consisting of multiple ICMEs and CIRs and that the shock was considerably accelerated by ICMEs. Similarly, the arrival times for the third and the fourth shocks in Model 2 are also delayed by $\sim$200 days compared to Model 1. We also note that the modest increase in $|\textbf{B}|$ of Model 1 around 2013.600 is not present in Model 2, supporting our view that it was affected by inclusion of all ICMEs in the OMNI data, some of which were directed far away from \textit{V1} in reality.

\section{Summary and Discussion}

Using daily averaged SW parameters from OMNI data as time-varying boundary conditions, we performed a global 3D time-dependent simulation to reproduce shocks propagating beyond the heliopause. The modeled shock arrival times closely match those of the late-2012 and mid-2014 shocks observed by \textit{V1}, though the changes in the model $|\textbf{B}|$ and density across the shocks are slightly larger than observed, considering the uncertainties in the measurements. Furthermore, the model predicts shock arrivals at \textit{V1} around 2017.395 and 2019.502. A variant of the model which excludes ICMEs from OMNI data suggests that ICMEs may accelerate some of the shocks significantly.  

Although we employed a reasonably high spatial resolution using multiple levels of AMR, the model did not reproduce the relatively steep drops in $|\textbf{B}|$ at 2013.353 and 2015.372, or the quasi-periodic oscillations of ISMF after the shock passage in mid-2014. We note that the radial grid of this model is small enough to reproduce daily fluctuations associated with solar activity well into the inner heliosheath. However, the non-radial grid size becomes too large to resolve small scale fluctuations (e.g., $\sim$28 days) deeper in the inner heliosheath where the flow develops significant non-radial components as the SW is diverted toward the tail, even with multiple levels of AMR. The computational cost for resolving these fluctuations would have been too excessive. A more detailed investigation of this phenomenon using sufficiently fine Cartesian AMR grids will follow.



\acknowledgments

The authors acknowledge use of the SPDF COHOWeb database and WSO data (\url{wso.stanford.edu}). This work is supported by the NSF PRAC award OCI-1144120 and related computer resources from the Blue Waters sustained-petascale computing project. Supercomputer time allocations were also provided on SGI Pleiades by NASA High-End Computing Program award SMD-15-5860 and on Stampede and Comet by NSF XSEDE project MCA07S033. TKK and NVP acknowledge support from the NSF SHINE project AGS-1358386, SAO subcontract SV4-84017, and NASA contract NNX14AF41G. LFB was supported by NASA contract NNG14PN24P.




\end{document}